\documentclass[prl,aps,reprint]{revtex4-1}
\usepackage{hyperref}
\usepackage{graphicx}
\usepackage{amsmath}

\DeclareMathOperator*{\argmin}{arg\,min}

\begin{document}

\title{Test of the Gravitational Redshift with {\sl Galileo} Satellites in an Eccentric Orbit}
\date{\today}
\author{Sven Herrmann$^1$, Felix Finke$^1$,  Martin L\"ulf$^2$, Olga Kichakova$^1$, Dirk Puetzfeld$^1$, Daniela Knickmann$^3$, Meike List$^1$, Benny Rievers$^1$, Gabriele Giorgi$^4$, Christoph G\"unther$^{2,4}$, Hansj\"org Dittus$^5$, Roberto Prieto-Cerdeira$^6$, Florian Dilssner$^7$, Francisco Gonzalez$^6$, Erik Sch\"onemann$^7$,  Javier Ventura-Traveset$^8$, Claus L\"ammerzahl$^1$}
\affiliation{$^1$University of Bremen, ZARM Center of Applied Space Technology and Microgravity, Bremen 28359, Germany}
\affiliation{$^2$Technical University Munich, Munich 80333, Germany}
\affiliation{$^3$OHB System AG, Bremen 28359, Germany}
\affiliation{$^4$Deutsches Zentrum f\"ur Luft- und Raumfahrt, Oberpfaffenhofen-Wessling 82234, Germany}
\affiliation{$^5$Deutsches Zentrum f\"ur Luft- und Raumfahrt, K\"oln 51147, Germany}
\affiliation{$^6$European Space and Technology Centre, ESA ESTEC, AZ Noordwijk 2201, Netherlands}
\affiliation{$^7$European Space Operations Centre, ESA ESOC, Darmstadt 64293, Germany}
\affiliation{$^8$European Space and Astronomy Centre, ESA ESAC, Villanueva de la Ca\~{n}ada, Madrid 28692, Spain}

\begin{abstract} 
On August 22, 2014, the satellites GSAT-0201 and GSAT-0202 of the European GNSS {\sl Galileo} were unintentionally launched into eccentric orbits. Unexpectedly, this has become a fortunate scientific opportunity since the onboard hydrogen masers allow for a sensitive test of the redshift predicted by the theory of general relativity. In the present Letter we describe an analysis of approximately three years of data from these satellites including three different clocks. For one of these we determine the test parameter quantifying a potential violation of the combined effects of the gravitational redshift and the relativistic Doppler shift. The uncertainty of our result is reduced by more than a factor 4 as compared to the values of Gravity Probe A obtained in 1976.
\end{abstract}

\maketitle

\section{Introduction}

The frequency shift that clocks experience in a gravitational potential is a central prediction of the theory of general relativity. The universal validity of this gravitational redshift is a consequence of the Einstein equivalence principle, which provides the underlying foundation of the theory \cite{Will2014}. The MICROSCOPE mission \cite{Touboul2017} could recently achieve an improved test of another aspect of the Einstein equivalence principle by testing the universality of free fall of two test masses in orbit with a precision of $\eta < 2 \times 10^{-14}$. The gravitational redshift has so far been tested with a much lower precision \cite{Pound1960,Pound1965,Vessot1979,Vessot1980}. Its first experimental verification was provided by Pound and Rebka in 1960 \cite{Pound1960}, who observed the shift using a M\"ossbauer emitter and absorber over a height difference of $\approx 23$\,m. The most accurate test so far was obtained by the Gravity Probe A (GPA) mission, which launched a hydrogen maser on board a sounding rocket to a height of 10\,000\,km above ground. During the flight the frequency generated by the maser on the rocket was compared with a corresponding maser on the ground. The  total relativistic frequency shift was found to be within $7 \times 10^{-5}$ of the value predicted by general relativity \cite{Vessot1979}.

In August 2014, an unexpected opportunity to reduce the uncertainty of this fundamental test even further arose through a problem during the launch of the satellites GSAT-0201 and GSAT-0202. They were erroneously placed into elliptical orbits. Today the eccentricity of the orbits is 0.16. Since these satellites carry passive hydrogen masers, this makes them attractive for a test of the gravitational redshift. Thus, in this letter we examine if and how the clock data from these satellites can, indeed, be used to reduce the uncertainty of such a test as compared to GPA. 

\section{Gravitational redshift for Galileo satellites}

The elapsed coordinate time $t$ of a clock moving in a weak gravitational potential $U$ at a velocity $v$ can be derived from 
\begin{equation}
\int dt =\int d\tau \left( 1-\frac{U}{c^2} + \frac{v^2}{2c^2} \right), \label{basic}
\end{equation}
where $t,$ and $\tau$ are the coordinate time and the clock's proper time, respectively. The second term in the bracket accounts for the gravitational redshift and the third for the relativistic Doppler effect. The associated time delay is routinely taken into account in global navigation satellite system (GNSS) receivers \cite{Ashby2003}. For a clock on a Kepler orbit of semimajor axis $a$ and eccentricity $e$ it can be written as  
\begin{equation}
\Delta t = \left( \frac{3GM_E}{2ac^2} + \frac{\Phi_0}{c^2} \right) t + \frac{2 \sqrt{G M_E a}}{c^2} e \sin E(t), \label{IGSredshift}
\end{equation}
with $E(t)$ being the eccentric anomaly of the orbit and $\Phi_0$ the gravitational potential at the location of the ground based reference clock. This equation includes a linear first term, and a second term that is modulated due to the eccentricity of the orbit. The latter is applied as a correction in GNSS receivers, typically reformulated in the form 
\begin{equation}
t_{\rm rel} = \frac{2\vec{v}\cdot\vec{r}}{c^2}. \label{IGSredshift2}
\end{equation}

With the {\sl Galileo} satellites GSAT-0201 and GSAT-0202, the eccentricity of the orbit is $e \approx 0.16$, and the relativistic eccentricity correction reaches a peak amplitude of approximately 370\,ns. This corresponds to a peak to peak modulation of the relative frequency of $\frac{\Delta f}{f} \approx 1\times10^{-10}$. A periodic modulation of this size is clearly discernible, given the relative frequency stability of the passive hydrogen maser clocks on-board. Ground-based relative frequency stability tests performed on these clocks show a flicker noise floor at the level of an Allan deviation of $\sigma \approx 10^{-14}$ at the time scale of an orbital revolution period T = 12.94\,h \cite{Rochat2012}.

\section{Data processing and analysis}

The pseudorange and carrier-phase observations from GSAT-0201 and GSAT-0202 have been collected by stations of the International GNSS Service Multi-GNSS Experiment network across the globe. This allows the estimation of the position $r$ and clock offsets $\tau$ for each satellite. For the purpose of the present experiment, this was specifically done by the Navigation Support Office of ESA ESOC. This ensured high precision, availability, reliability, and full control of the process.

The clock data are sampled every 30\,s interval, the satellite position and velocity are estimated and provided every 300\,s from which intermediate values are obtained by interpolation.  The clock estimates feature random walk noise and a linear drift, which is on the order of $1\,\mu$s/h and results from a residual contribution of the first term of equation (\ref{IGSredshift}) as well as the natural drifts of the satellite clock and the clock of the selected reference ground station.

The data considered in our analysis starts on January 11, 2015, extending to December 16, 2017, thus, covering almost three years. Every {\sl Galileo} satellite carries two rubidium and two hydrogen maser clocks, with only one of them being the active transmission clock, as configured by operations. During our period of observation, the data were obtained from five different clocks on two satellites labeled clock 1 to 5 as given in Table \ref{TableClocks}. Clock 4 is a Rubidium Atomic Frequency Standard (RAFS) on GSAT-0202 which is of inferior clock stability as compared to the passive hydrogen maser (PHM) clocks. Clock 2 is linked to PHM A on GSAT-0201, for which ESA has confirmed non-nominal drift during this period. Thus, we do not include these two clocks in our further analysis and focus on the data from the nominal PHM clocks 1, 3 and 5 only. 

\begin{figure}
\includegraphics[width=\columnwidth]{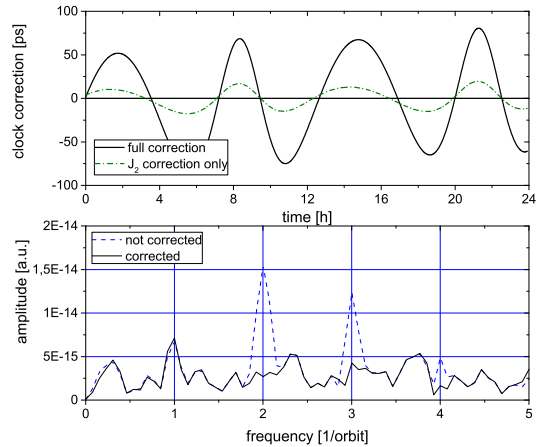}
\caption{Top: Correction as applied to the ESOC clock data for the first day in GPS week 1938 (a linear drift has been removed). The correction is the difference of Eq. (\ref{IGSredshift2}) and the numerically integrated redshift model along the perturbed real orbit of GSAT-0202. This includes the redshift contribution from the mass quadrupole ($J_2$) which is shown separately in addition. Bottom: An FFT of the differentiated clock estimates of GSAT-0202 for GPS week 1938 before and after the correction was applied.}
\label{fig1}
\end{figure}

Preprocessing of the clock data as done by the Navigation Support Office of ESA ESOC already included the relativistic correction given in Eq. (\ref{IGSredshift2}). This model, however, is not sufficient to meet the requirement of a precision test, as the equation is derived assuming an ideal Kepler orbit without perturbations. Thus, we first removed the correction of Eq. (\ref{IGSredshift2}) and applied a refined relativistic correction instead. This refined correction is obtained by numerically integrating Eq. (\ref{basic}) along the orbit and by including the quadrupole moment into Earth's gravitational potential $U$. The latter moment is not included in the clock estimates as provided by ESOC. Correspondingly, a Fourier analysis of the raw clock data shows a peak at twice the orbital revolution frequency. This peak is significantly decreased when the improved corrections are applied as shown in Fig. \ref{fig1}. 

\begin{table}
\begin{ruledtabular}
\begin{tabular}{ccccc}
Clock/Sat & Start & End & Span [d] & Clock no. \\
 \hline PHM-B/0201 & 2015/01/11  & 2016/06/15 & 522 & 1 \\
 PHM-A/0201 & 2016/07/02 & 2017/12/16 & 533 & 2 \\
 PHM-B/0202 & 2015/03/19 & 2015/11/04 & 231 & 3 \\
 RAFS/0202 & 2015/11/05 & 2016/07/02 & 241 & 4  \\
 PHM-A/0202 & 2016/07/03 & 2017/12/16 & 532 & 5 \\
\end{tabular}
\end{ruledtabular}
\caption{Clocks set for transmission on the respective satellites during the measurement span.}\label{TableClocks}
\end{table}

Before entering into the analysis outliers have been identified and removed from the clock estimates based on a $5 \sigma$ criterion, typically removing only a few data points per day. In addition few days ($<3$ per clock) with strongly disturbed data have been removed upon manual inspection.

To estimate the magnitude of a possible violation of general relativity we introduce a test parameter $\alpha$ that quantifies any deviation from the refined relativistic model, following the definition given in \cite{Will2014}. There, a violation of the relative frequency shift of a stationary clock due to a change in the gravitational potential $\Delta U$ is modeled as
\begin{equation}
\frac{\Delta f}{f} = (1+\alpha_{rs})\frac{\Delta U}{c^2}
\label{basic2}
\end{equation}
where $\alpha_{rs} = 0$ complies with the prediction of general relativity. A corresponding test parameter $\alpha$ including the relativistic Doppler effect is used in our model and is determined in a least squares fit to the postprocessed clock estimates. We explored two options for this fit. The first one is the application of a model based on numerical integration of Eq. (\ref{basic}). The second one is to numerically differentiate the clock data $\tau$, instead, and to use the integrand of Eq. (\ref{basic}) as our model, which describes the relative frequency shift rather than the elapsed time. The latter approach has the advantage in that it converts the predominant random walk noise into white noise. This leaves only small residual correlations in the fitted data, which we take into account by using a generalized least squares algorithm \cite{FGLS, Baltagi}. 

The full relativistic model in this case is given by 
\begin{equation}
f_{rel,i} = -\frac{G M_E}{r_i c^2}\left[ 1- \frac{J_2 a^2_E}{r_i^2}\left( \frac{z_i^2}{r_i^2}-\frac{1}{2}\right) \right)] +\frac{v_i^2}{2c^2} ,
\label{model}
\end{equation}  
where $J_2$ describes Earth's mass quadrupole, $a_E$ is Earth's equatorial radius, and $r_i$, $z_i$ and $v_i$ are provided by the orbit solution with $i$ indexing the samples in the time series of data. The latter have been transformed into an inertial frame and interpolated at the time of the clock samples. This model is then scaled by the test parameter $\alpha$ and fitted to the postprocessed and differentiated clock data $f_i$ for each day separately. The clock is modeled by a constant offset $a_{1}$ for each day. The latter time span is used for the fit, since orbits are processed day by day without continuity requirements at daily boundaries. Thus, the estimates of $\alpha$ and $a_1$ are obtained from the equation 
\begin{equation}
(\hat{\alpha}, \hat{a}_1) = \argmin\sum_i^N \left( f_i - \alpha f_{rel,i} - a_{1}  \right)^2,
\end{equation}
with a maximum of $N = 2880$ samples per day. 

To validate our method we analyzed two types of test data sets where we either injected an artificial test signal with $\alpha \neq 0$ based on our model (\ref{model}) into the original data or superimposed a test signal with modeled drift and random walk noise. This way, we could verify that the mean value and error estimate of our final result are consistent and potential error sources, e.g., from numerical differentiation due to the finite sample size or residual correlation of fit parameters are negligible throughout the analysis.

\begin{figure}
\includegraphics[width=\columnwidth]{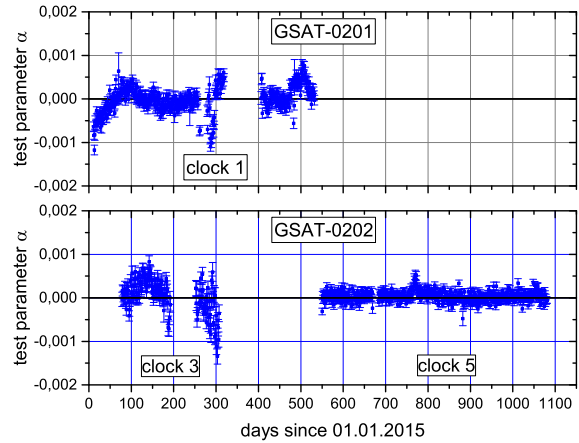}
\caption{Distribution of daily test parameter results for clocks 1, 3 and 5. Error bars are the $1\sigma$ fit errors from the respective least squares fit. Gaps in the data of clocks 1 and 3 are when there was no clock transmission. Note, also, that we exclude 13 days in the data of clock 5 coinciding with an interruption of clock operation of PHM A following day 670.}
\label{fig2}
\end{figure}

Upon processing the data of all three clocks we are left with a distribution of fitted test parameters $\alpha$ for the three clocks as shown in Fig. \ref{fig2}. Then, we determine the weighted average of the daily $\alpha$ results for each clock using the inverse squared fit error as a weight, as well as the combined statistical uncertainty. From this we obtain the results given in the third line of Table \ref{errorbudget}. 
The (varying) bias observed in the distribution of all three clocks indicates that there is one or several systematic effects present that need to be addressed carefully. Thus, before discussing the distributions further, we continue with an assessment of the direct effects onto the clock frequency from magnetic fields and temperature variations, as well as systematic model uncertainties.

\section{Estimate of systematic uncertainties}

First, we consider a potential systematic error in the applied orbit solution. Here, any radial error in the orbit estimation will map into the clock estimate as $\delta \tau_i = \delta r_i/c$ potentially introducing a systematic error. Imperfect modeling of the reaction of the satellite to solar radiation pressure is a potential cause of such systematic orbit errors. This may reach a size of a few centimeters, see, e.g., \cite{Montenbruck2015}. To assess the magnitude of such a systematic orbit model error, additional and independent laser ranging measurements by the International Laser Ranging Service (ILRS) are used. They are made possible by the retroreflectors mounted on each of the satellites. A dedicated effort was made by ILRS to support the project by increasing the number of measurements for the satellites GSAT-0201 and 0202, see \cite{ILRS2016}. From these laser ranging measurements, the radial one-way residuals with respect to the modeled orbit solution were generated by ESOC. These can be converted into a bias in the time domain by dividing by $-c$. 

Then, in a first approach, we fit these biases with our redshift model on a daily basis following the same approach that was applied to the clock data. This results in a distribution of daily $\alpha$ corrections. Then, from this, we derive the correction to the $\alpha$ mean value and its associated uncertainty. The accuracy of this estimate, however, is limited because there are only few satellite laser ranging (SLR) measurements per day which increases the uncertainty of each daily fit. Thus, as an alternative approach, we also do a global linear regression of the converted SLR data and the corresponding redshift model data for each clock. We, thereby, observe a good agreement from the results of both methods in our estimates for clock 5, while, for clocks 1 and 3, we only rely on the result from the global regression.

Next, we address the effect of varying magnetic fields. Since the {\sl Galileo} satellites are not equipped with magnetometers, we relied on the International Geomagnetic Reference Field model for Earth's magnetic field \cite{Thebault2015}. The expected variation of the magnetic field due to the satellites' movement is of the order of few mG, typically spanning $\approx 7$\,mG peak to peak. Consulting data from the THEMIS mission \cite{THEMIS2018} as well as the Tsyganenko model \cite{Tsyganenko2007} at {\sl Galileo} altitudes, we estimate the uncertainty of our model to be less than 10\%. The effect on the clock was estimated using the sensitivity of the PHMs to be $\frac{\Delta f}{f} = 3\times 10^{-13}$/G as given in \cite{Rochat2012}. We took this to be the sensitivity for a field applied along the most sensitive clock axis, since no further information on field orientation is given in \cite{Rochat2012}. A conservative estimate was obtained by assuming that the effect is caused by the total magnetic field rather than by its projection onto a specific PHM axis. Using this model we correct the clock estimates and rerun the analysis as described above resulting in a corrected $\alpha$ estimate. Doing so we consider both added ($\alpha_+$) or subtracted ($\alpha_-$) correction, since no information is available from \cite{Rochat2012} to deduce the sign of the imposed frequency shift. Then, for the final result, we use the difference $\Delta \alpha_B = \alpha_+ - \alpha_-$ as the total uncertainty from magnetic field variations.

The effect of temperature variations is also estimated using the sensitivity given in \cite{Rochat2012} which is $\frac{\Delta f}{f} = 2\times 10^{-14}$/K. The clocks on board the satellites are actively temperature stabilized within a maximum allowed deviation of $\Delta T = \pm 0.5$\,K. The thermal control period is of the order of 10\,min and, thus, more than a factor of 50 faster than the period of orbital revolution \cite{PrivateComm}. Thus averaging over many control oscillations per orbit should already significantly suppress the effective temperature variation at orbital frequency. In addition, the incident solar radiation presents the dominant source of temperature variation and the phase of any residual temperature systematic at orbital frequency should be fixed relative to the sun position. Thus further decorrelation with a redshift violation signal occurs when analyzing data spanning more than one year. Based on this, we conservatively assume a combined suppression factor of 10 equivalent to a maximum contribution from temperature variation to $\alpha$ of $\pm 2 \times 10^{-5}$ in each clock. 

In addition to the effects discussed above, further possible systematics from reference ground clocks, atmospheric corrections or phase wind up due to the rotation of the satellite about the earth-pointing axis have been assessed. We estimate their systematic contribution to $\alpha$ at the low $10^{-6}$ level, since they are averaged over many ground stations and a possible diurnal modulation will be disentangled from a redshift violation signal at orbital period within a few days of measurement.

\begin{figure}
\includegraphics[width=\columnwidth]{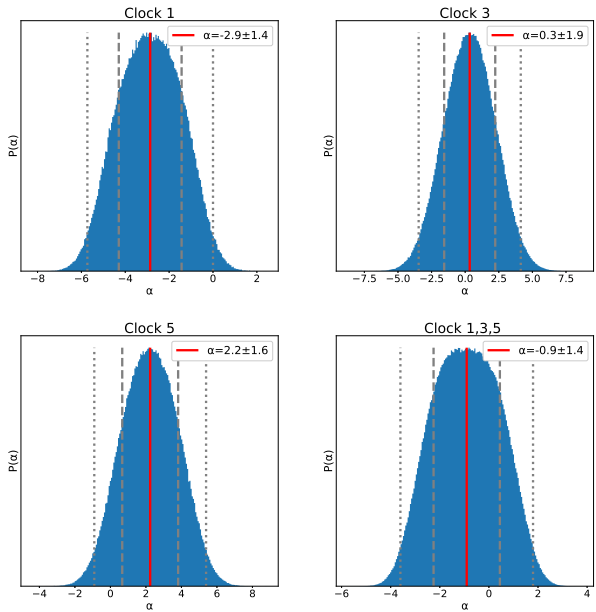}
\caption{Shown are the marginalized posteriors including the systematic corrections for the single clocks as well as for the combination of all three clocks. The vertical lines in the plots mark the posterior mean as well as the equal tailed 68\% (dashed lines) and 95\% (dotted lines) intervals. Numbers stated in the graphs are all times $10^{-5}$. }
\label{fig5}
\end{figure}

All contributions entering our final result are summarized in Table \ref{errorbudget}. To combine all systematic corrections and their uncertainties we follow a basic Bayesian approach \cite{Gelman2013}. For each clock we derive the final mean value and its uncertainty from the marginalized posterior including the different systematic corrections. Orbit corrections were modeled by a normally distributed prior, while for the temperature and magnetic field corrections we assumed uniform distributions in the ranges given in Table \ref{errorbudget}. The final uncertainty we state on each clock is the equal tailed 68$\%$ interval (see Fig. \ref{fig5}).


\begin{table}
\begin{ruledtabular}
\begin{tabular}{cccc}
$\alpha \pm \Delta \alpha [\times 10^{-5}]$  &  Clock 1 &  Clock 3 & Clock 5 \\
\hline Days included   &  414 & 167 & 510 \\
 \hline Statistics & $-0.7 \pm 0.5$ & $8.4 \pm 1.2$ & $3.7 \pm 0.4$ \\
 \hline Orbit model & $-2.2 \pm 0.5 $ & $-8.1 \pm 0.9$ & $ -1.5 \pm 0.9$ \\
 Temperature & $0 \pm 2.0 $ & $0 \pm 2.0 $ & $0 \pm 2.0$ \\
 Magnetic field & $0 \pm 0.8$ & $0 \pm 0.8$ & $0 \pm 0.8$ \\
\hline Total & $-2.9 \pm 1.4$ & $0.3 \pm 1.9$ & $2.2 \pm 1.6$ \\
\end{tabular}
\end{ruledtabular}
\caption{Error budget of statistical and systematic uncertainties. Note that the different contributions refer to different underlying probability distributions. The statistical error is the $1\sigma$ error as determined from the distribution of fit results. Systematic uncertainties for magnetic fields and temperature are upper and lower bounds (assuming uniform distribution). The orbit uncertainty is, again, a 1$\sigma$ interval. The total result for each clock is then derived from the posterior of the combined corrections and the stated uncertainty represents the equal tailed 68$\%$ interval (see Fig. \ref{fig5}).}\label{errorbudget}
\end{table}

\section{Discussion}

The orbit correction derived from independent laser ranging data significantly reduces the observed bias for the two clocks on GSAT-0202. Further accounting for the uncertainties due to temperature and magnetic field variations, the results for all three clocks are eventually consistent with the GR prediction within approximately 2$\sigma$. Then, if we take the combined posterior for all three clocks, we are left with a posterior mean of $\alpha = (-0.9 \pm 1.4)\times 10^{-5}$, which would correspond to a fivefold improvement over GPA. 

A closer look at the distributions of daily results for clocks 1 and 3 in Fig. \ref{fig2}, however, reveals rather pronounced slow variations of the bias. While these average out to a certain extent in the total result, their magnitude clearly exceeds our boundaries on daily temperature and magnetic field effects. Also, for these clocks we are unable to correlate the observed variations with the orbit model error, partially due to the low number of SLR measurements per day. 

For clock 5 on the other hand, we have no evidence for an additional unaccounted systematic influence, apart from a short ''glitch'' between days 767 and 773. In addition, we get a consistent result from daily and global SLR fits to model the orbit correction for this clock. Thus, we consider the above bounding approach based on existing models and reasonable conservative assumptions to be sufficient to reliably estimate the effects in clock 5 and decide to take our final result from this clock only. This leaves us with $\alpha = (2.2 \pm 1.6) \times 10^{-5}$, the uncertainty of which is still a factor of 4 below that of GPA. The apparent deviation from zero is less than 2$\sigma$ and is partially attributed to the observed ''glitch''. Removal of these seven days changes the result to $\alpha = (1.9 \pm 1.6) \times 10^{-5}$. Ongoing improvements in the processing, in particular in the modeling of the clocks, might lead to tightened results in the future.\\

The above analysis provides a test of the combined effect of gravitational redshift and relativistic Doppler effect, similar as done in the analysis of GPA given in \cite{Vessot1979}, and, thus, allows for a direct comparison to the result given there. The relativistic Doppler effect, however, has been tested separately and at significantly better precision in other experiments already \cite{Botermann2014}. If we restrict our violation model only to the gravitational redshift part as in equation (\ref{basic2}) we obtain a combined result of $\alpha_{rs} = (4.5 \pm 3.1)\times 10^{-5}$, which, again, provides a fourfold reduced uncertainty as compared to $\alpha_{rs}< 1.4 \times 10^{-4}$ as given for GPA in \cite{Vessot1989}.

As data taking of the GSAT-0201 and 0202 satellites continues, prospects for further improvements are limited by the uncertainty in temperature and magnetic field systematics due to the lack of sensors and telemetry. These limitations could be overcome by direct measures if a similar but dedicated mission was done in the future, which could be an interesting complement to other upcoming precision tests of the gravitational redshift, e.g., from the ACES mission \cite{Meynadier2018} or the RadioAstron mission \cite{Litvinov2018}.\\

\begin{acknowledgments}

We acknowledge support from the European Space Agency and the General Studies Program for support of the GREAT Project as well as from the ESA navigation support office at ESOC. We further acknowledge funding from the DLR Space Administration with funds provided by the Federal Ministry of Economic Affairs and Energy (BMWi) under Grant No. 50WM1547 and No. 50WM1548. C.L., D.P., B.R., and M. List acknowledge support from the Deutsche Forschungsgemeinschaft (DFG) through the Collaborative Research Center (SFB) 1128 geo-Q. D.P. further acknowledges support of the Deutsche Forschungsgemeinschaft (DFG) through Grant No. PU 461/1-1. We would also like to thank Pacome Delva, Tim Springer, Francesco Gini, Luigi Cacciapuoti, Pierre Waller, Thomas Driebe and Fritz Merkle for valuable discussions and continuous support.

\end{acknowledgments}

{\sl Note added in proof}. -- We note that in the framework of the GREAT project another team has conducted an independent analysis of the same data as considered in this Letter. Their results are published in the same issue \cite{Delva2015}.

\bibliography{Bibliography}

\end{document}